\documentclass[aps,prd,twocolumn,superscriptaddress,showpacs]{revtex4}
\usepackage{amsmath}
\usepackage{graphicx}

\newcommand{\bastar}{\begin{eqnarray*}}
\newcommand{\eastar}{\end{eqnarray*}}
\newskip\humongous \humongous=0pt plus 1000pt minus 1000pt

\newif\ifdtup

\relax

\newcommand{\bea}{\begin{eqnarray}}
\newcommand{\eea}{\end{eqnarray}}

\newcommand{\A}{{\vec A}}
\newcommand{\F}{{\vec F}}

\newcommand{\nn}{\nonumber}
\newcommand{\Int}{\displaystyle{\int}}
\begin{document}
\title{Electroweak Monopole Production at the LHC - a Snowmass White Paper}
\author{Y. M. Cho}
\email{ymcho7@konkuk.ac.kr}
\affiliation{Administration Building 310-4, Konkuk University, 
Seoul 143-701, Korea}
\affiliation{School of Physics and Astronomy \\ 
Seoul National University, Seoul 151-747, Korea}
\author{James Pinfold}
\email{jpinfold@ualberta.ca}
\affiliation{Physics Department \\
University of Alberta, Edmonton, Alberta T6G 0V1, Canada.}

\begin{abstract}
We maintain that the  search for the electroweak monopole is  a key issue
in the  advancement of our understanding of the standard model. Unlike the Dirac monopole 
in electrodynamics, which is optional, the electroweak monopole 
should exist within the framework of the standard model. 
The mass of the electroweak monopole is estimated to be 5 to 7 TeV, 
but could be as large as 15 TeV. Above threshold its production rate 
at  the LHC is expected to be relatively large, $(1/\alpha_{em})^2$ 
times bigger than that of W$^{+}$W$^{-}$ pairs. 
The search for the electroweak monopole is one of the prime 
motivations of the newest LHC experiment, MoEDAL, which is due to start 
data taking in 2015.
\end{abstract}
\maketitle
\section{Introduction}

A new particle has recently been discovered at the LHC by 
the  ATLAS and CMS  experiments \cite{LHC}.  As more data is analyzed this new particle
looks  increasingly like the Standard Model Higgs boson. If indeed the Standard Model
Higgs boson has been discovered conventional wisdom tells us  that this  is the 
final crucial test of the Standard Model. However, we emphasize that there is another 
fundamental entity that  should  arise from the framework of the Standard Model
- this is the Electroweak (EW), or ``Cho-Maison'', magnetic monopole  \cite{plb97,yang}. 
We maintain that the search for the EW monopole is of key importance in advancing 
our understanding of the  Standard Model. 

What is the genesis of the Cho-Maison monopole?  In electrodynamics the $U(1)$ gauge group 
need not be non-trivial, so that  Maxwell's theory does not have 
to have a monopole. Only when the gauge group $U(1)$ becomes 
non-trivial do we have Dirac's monopole. In the standard model, 
however, the gauge group is $SU(2)\times U(1)_Y$. The electromagnetic 
$U(1)_{em}$ comes from the $U(1)$ subgroup of $SU(2)$ and the 
hypercharge $U(1)_Y$. But it is well known that the $U(1)$ subgroup 
of $SU(2)$ is non-trivial, due to the non-Abelian nature. This 
automatically makes the $U(1)_{em}$ non-trivial, so that the standard 
model should have an electroweak monopole \cite{plb97,yang}. So, if the 
standard model is correct, the Cho-Maison monopole must exist. 

It has been asserted that the Weinberg-Salam model has no 
topological monopole of physical interest \cite{vach}. 
The basis for this ``non-existence theorem" is that with 
the spontaneous symmetry breaking the quotient space 
$SU(2) \times U(1)_Y/U(1)_{em}$ allows no non-trivial second 
homotopy. This claim, however, is unfounded.

Actually the Weinberg-Salam model, with the hypercharge 
$U(1)$, could be viewed as a gauged $CP^1$ model in which 
the (normalized) Higgs doublet plays the role of the $CP^1$ 
field. So the Weinberg-Salam model does have exactly 
the same nontrivial second homotopy as the Georgi-Glashow model 
which allows the 'tHooft-Polyakov monopole \cite{plb97}. 

The Cho-Maison monopole is the electroweak generalization of 
the Dirac's monopole, so that it could be viewed as a hybrid 
of Dirac and 'tHooft-Polyakov monopoles. But unlike the Dirac's 
monopole, it carries the magnetic charge $(4\pi)/e$. This is
because in the standard model the $U(1)_{em}$ has the period 
of $4\pi$, not $2\pi$, as it comes from the $U(1)$ subgroup of 
$SU(2)$. This makes thesingle  magnetic charge of the electroweak 
monopole twice as large as that of the Dirac  Monopole.   

\section{The Electroweak Monopole}

Consider the Weinberg-Salam model,
\begin{gather}
{\cal L} =-\frac{1}{4}\F_{\mu\nu}^2 
 -\frac{1}{4}G_{\mu\nu}^2 -|D_{\mu} \phi|^2 
  -\frac{\lambda}{2}\big(|\phi|^2
  -\frac{\mu^2}{\lambda}\big)^2 , \nonumber \\
D_{\mu} \phi = \big(\partial_{\mu} 
  -i\frac{g}{2} \vec \tau \cdot \A_{\mu}
  -i\frac{g'}{2} B_{\mu}\big) \phi,
\label{lag1}
\end{gather}
where $\phi$ is the Higgs doublet, $\F_{\mu\nu}$ and 
$G_{\mu\nu}$ are the gauge field strengths of $SU(2)$ 
and $U(1)_Y$ with the potentials $\A_{\mu}$ and $B_{\mu}$. 
Now choose the static spherically symmetric ansatz
\begin{gather}
\phi=\frac{1}{\sqrt{2}}\rho(r)\xi(\theta,\varphi),
~~~\xi =i\left(\begin{array}{cc}
\sin (\theta/2)\,\, e^{-i\varphi}\\
- \cos(\theta/2) \end{array} \right), \nn\\
\A_{\mu}= \frac{1}{g} A(r) \partial_{\mu} t~\hat r
+\frac{1}{g}(f(r)-1)~\hat r \times \partial_{\mu} \hat r, \nn\\
B_{\mu} =\frac{1}{g'} B(r) \partial_{\mu}t 
-\frac{1}{g'}(1-\cos\theta) \partial_{\mu} \varphi.  
\label{ans}
\end{gather}
To proceed notice that we can Abelianize (\ref{lag1}) gauge 
independently using the Abelian decomposition \cite{cho}. 
With the gauge independent Abelianization the Lagrangian is 
written in terms of the physical fields as
\bea
&{\cal L}= -\dfrac{1}{2}(\partial_\mu \rho)^2 
-\dfrac{\lambda}{8}\big(\rho^2-\rho_0^2 \big)^2 \nn\\
&-\dfrac{1}{4} {F_{\mu\nu}^{\rm (em)}}^2 
-\dfrac{1}{4} Z_{\mu\nu}^2-\dfrac{g^2}{4}\rho^2 |W_\mu|^2
-\dfrac{g^2+g'^2}{8} \rho^2 Z_\mu^2 \nn\\
&-\dfrac{1}{2}|(D_\mu^{\rm (em)} W_\nu - D_\nu^{\rm (em)} W_\mu)
+ ie \dfrac{g}{g'} (Z_\mu W_\nu - Z_\nu W_\mu)|^2  \nn\\
&+ie F_{\mu\nu}^{\rm (em)} W_\mu^* W_\nu
+ie \dfrac{g}{g'}  Z_{\mu\nu} W_\mu^* W_\nu \nn\\
&+ \dfrac{g^2}{4}(W_\mu^* W_\nu - W_\nu^* W_\mu)^2,
\label{lag2}
\eea
where $\rho$, $W_\mu$, $Z_\mu$ are the Higgs, $W$, $Z$ bosons, 
$D_\mu^{\rm (em)}=\partial_\mu+ieA_\mu^{\rm (em)}$, and 
$e=gg'/\sqrt{g^2+g'^2}$ is the electric charge. 

Moreover, the ansatz (\ref{ans}) becomes 
\bea
&\rho =\rho(r),~~~W_\mu= \dfrac{i}{g}\dfrac{f(r)}{\sqrt2}e^{i\varphi}
(\partial_\mu \theta +i \sin\theta \partial_\mu \varphi),
\nn\\
&A_\mu^{(em)}=e\Big(\dfrac{A(r)}{g^2}+\dfrac{B(r)}{g'^2} \Big)
\partial_\mu t -\dfrac1e (1-\cos\theta)\partial_\mu \varphi,  \nn\\
&Z_\mu= \dfrac{e}{gg'}\big(A(r)-B(r) \big)\partial_\mu t.
\label{ansatz}
\eea
With this we have the following equations of motion
\begin{gather}
\ddot{\rho}+\frac{2}{r} \dot{\rho}-\frac{f^2}{2r^2}\rho
=-\frac{1}{4}(A-B)^2\rho +\frac {\lambda}{2}\big(\rho^2
-\frac{2\mu^2}{\lambda}\big)\rho , \nn\\
\ddot{f}-\frac{f^2-1}{r^2}f=\big(\frac{g^2}{4}\rho^2
- A^2\big)f, \nn\\
\ddot{A}+\frac{2}{r}\dot{A}-\frac{2f^2}{r^2}A
=\frac{g^2}{4}\rho^2(A-B), \nn \\
\ddot{B} +\frac{2}{r} \dot{B}
=-\frac{g'^2}{4} \rho^2 (A-B).
\label{cmeq} 
\end{gather}
Obviously this has a trivial solution
\bea
\rho=\rho_0=\sqrt{2\mu^2/\lambda},~~~f=0,
~~~A=B=0,
\eea
which describes the point monopole in Weinberg-Salam model
\bea
A_\mu^{\rm (em)}=-\frac{1}{e}(1-\cos \theta) \partial_\mu \varphi.
\eea
This monopole has two remarkable features. First, this is the 
electroweak generalization of the Dirac's monopole, but not 
the Dirac's monopole. It has the electric charge $4\pi/e$, not 
$2\pi/e$ \cite{plb97}. Second, this monople naturally admits a 
non-trivial dressing of weak bosons. With the non-trivial 
dressing, the monopole becomes the Cho-Maison dyon.

Indeed with the boundary condition
\bea
&\rho(0)=0,~~f(0)=1,~~A(0)=0,~~B(0)=b_0, \nn\\
&\rho(\infty)=\rho_0,~f(\infty)=0,~A(\infty)=B(\infty)=A_0,
\label{bc0}
\eea
we can show that the equation (\ref{cmeq}) admits a family of 
solutions labeled by the real parameter $A_0$ lying in the 
range \cite{plb97,yang} 
\bea
0 \leq A_0 < {\rm min} ~\Big(e\rho_0,~\frac{g}{2}\rho_0\Big).
\label{boundA} 
\eea 
From this we have the electroweak dyon shown in Fig. 1, which 
becomes the Cho-Maison monopole when $A=B=0$. Since $A_\mu^{(em)}$ 
has the point monopole, the solution can be viewed as a singular 
monopole dressed by $W$ and $Z$ bosons. This confirms that it can 
be viewed as a hybrid of the Dirac monopole and the 'tHooft-Polyakov 
monopole (or Julia-Zee dyon in general). 

\begin{figure}
\includegraphics[width=9cm]{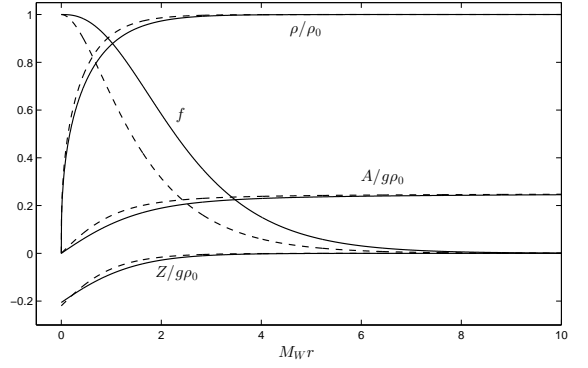}
\caption{The finite energy electroweak dyon solution. 
The solid line represents the finite energy dyon 
and dotted line represents the Cho-Maison dyon, where 
$Z=A-B$ and we have chosen $f(0)=1$ and $A(\infty)=M_W/2$.}
\label{fig2}
\end{figure}

To find the monopole experimentally it is important 
to estimate its  mass. At the classical level it carries an infinite 
energy because of the point singularity at the center, but from 
the physical point of view it must have a finite energy. To 
estimate the mass let
\bea
&K_A =  \dfrac{1}{4} \Int\vec{F}_{ij}^2 d^3x,
~~~K_B=\dfrac{1}{4} \Int B_{ij}^2 d^3x  \nn\\
&K_\phi=\Int |{\cal D}_i \phi|^2 d^3 x,
~V_\phi= \dfrac{\lambda}{2} \Int\big(|\phi|^2
-\dfrac{\mu^2}{\lambda} \big)^2 d^3x,
\eea
and divide the energy to infinite and finite parts
\bea
&E=E_0+E_1,  \nn\\
&E_0=K_B,~~~~E_1= K_A + K_\phi + V_\phi.
\eea
With $A=B=0$ we have 
\bea
&K_A= \dfrac{4\pi}{g^2} \Int_0^\infty 
\big(\dot{f}^2 + \dfrac{(f^2-1)^2}{2r^2} \big) dr, \nn\\
&K_B=\dfrac{2\pi}{g'^2}\Int_0^\infty \dfrac{1}{r^2}dr,
~~~K_\phi= 2\pi \Int_0^\infty (r\dot{\rho})^2 dr,  \nn \\
&V_\phi=\dfrac{\pi}{2} \Int_0^\infty \lambda r^2 \big(\rho^2 
-\rho_0^2 \big)^2 dr.
\eea
Clearly $K_B$ makes the monopole energy infinite. So we have to 
regularize it to make the monopole energy finite.

Suppose an ultra-violet regularization coming from quantum 
correction makes $B_K$ finite. Now, under the scale transformation 
\bea
\vec x \rightarrow \lambda \vec x,
\eea 
we have  
\bea
&K_A \rightarrow \lambda K_A,~~~K_B \rightarrow \lambda K_B, \nn\\
&K_\phi \rightarrow\lambda^{-1} K_\phi,  
~~~V_\phi \rightarrow \lambda^{-3} V_\phi.
\eea
So we have the following energy minimization condition for 
the stable monopole 
\bea
K_A + K_B = K_\phi + 3V_\phi.
\label{derrick}
\eea
From this we can infer the value of $K_B$. For the Cho-Maison 
monopole we have (with $M_W \simeq 80.4~{\rm GeV}$, 
$M_H\simeq 125~{\rm GeV}$, and $\sin^2\theta_{\rm w}=0.2312$) 
\bea
&K_A \simeq 0.1852 \times\dfrac{4\pi}{e^2}{M_W},
~~~~K_\phi \simeq 0.1577 \times\dfrac{4\pi}{e^2}{M_W},  \nn \\
&V_\phi \simeq 0.0011 \times\dfrac{4\pi}{e^2}{M_W}.
\eea
This, with (\ref{derrick}), tells that
\bea
&K_B \simeq 0.0058 \times \dfrac{4\pi}{e^2} M_W,  \nn\\
&E \simeq 0.3498 \times \dfrac{4\pi}{e^2} M_W \simeq 3.85~{\rm TeV}.
\eea
This strongly implies that the electroweak monopole of mass 
around 4 TeV could be possible \cite{cho}. 

\begin{figure}
\includegraphics[width=8.5cm]{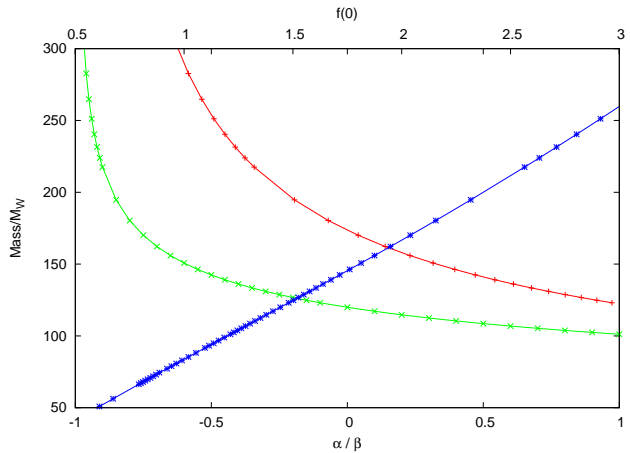}
\caption{The energy dependence of the electroweak monopole
on $\alpha$, $\beta$, or $f(0)$. The red and green curves 
represents the $\alpha$ and $\beta$ dependence, and the blue 
curve represents the $f(0)$ dependence.}
\label{fig3}
\end{figure}

To backup the above argument, suppose the quantum correction 
induces the following modification of (\ref{lag2})
\begin{gather}
\delta {\cal L}=ie \alpha F_{\mu\nu}^{\rm (em)} W_\mu^* W_\nu 
+\beta \frac{g^2}{4}(W_\mu^*W_\nu-W_\nu^*W_\mu)^2,  
\end{gather}
where $\alpha$ and $\beta$ are the quantum correction of the 
coupling constants. With this we can make the monopole energy 
finite with 
\begin{gather}
\label{cond1}
f^2(0)=\frac{1+\alpha}{1+\beta},
~~~~~\frac{(1+\alpha)^2}{1+\beta}=\frac{g^2}{e^2}.
\end{gather}
So only one of the three parameters $\alpha$, $\beta$, $f(0)$, 
becomes arbitrary.  Now, with $f(0)=1$, we have the finite 
energy monopole with energy $E\simeq  6.72~{\rm TeV}$. This is 
shown in Fig. 1. In general the energy of the monopole depends 
on the parameter $f(0)$, $\alpha$, or $\beta$, and this dependence 
is shown in Fig. 2. 

This strongly supports our prediction of the monopole mass 
based on the scaling argument. Moreover, this confirms that 
a minor quantum correction could regularize the Cho-Maison 
monopole and make the energy finite \cite{cho}. 

Moreover, in the absence of the $Z$-boson (\ref{lag2})
reduces to the Georgi-Glashow Lagrangian when the coupling 
constant of the quartic self interaction and the mass of 
the $W$-boson change to $e^2/g^2$ and $(e^2/g^2) \rho_0$.
In this case (\ref{cmeq}) reduces to the following 
Bogomol'nyi-Prasad-Sommerfield equation in the limit 
$\lambda=0$ \cite{prasad}
\bea
\dot{\rho}\pm \dfrac{1}{er^2}\big(\dfrac{e^2}{g^2}f^2-1 \big)=0,  \nn\\
\dot{f}\pm e\rho f=0.
\label{self2}
\eea
This has the analytic monopole solution
\begin{gather}
\rho=\rho_0\coth(e\rho_0r)-\frac{1}{er},  \nn\\
f= \frac{g\rho_0 r}{\sinh(e\rho_0r)},
\end{gather}
whose energy is given by the Bogomol'nyi bound
\begin{eqnarray}
E=\frac{8\pi}{e^2}\sin \theta_{\rm w~}M_{W}
\simeq 5.08~{\rm TeV}.
\end{eqnarray}
The Cho-Maison monopole, the regularized monopole, and the 
analytic monopole are shown in Fig. 3. From this we can 
confidently say that the mass of the electroweak 
monopole could be around 4 to 7 TeV.

\begin{figure}
\includegraphics[width=9cm]{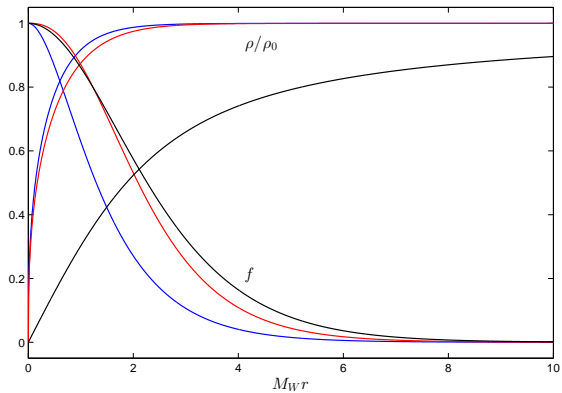}
\caption{The electroweak monopoles. The blue, red, and black 
curves represent the Cho-Maison monopole, the regularlized monopole, 
and the analytic monopole, respectively.}
\label{fig3}
\end{figure}

Independent of the details there is a simple argument which 
can justify the above estimate. Roughly speaking, the mass of 
the electroweak monopole should come from the same mechanism 
which generates the mass of the weak bosons, except that the 
coupling is given by the monopole charge. This means that the 
monopole mass should be of the order of $M_W/\alpha_{em}\simeq$ 
10.96 TeV, where $\alpha_{em}$ is the electromagnetic fine 
structure constant. This supports the above mass estimate \cite{cho}.

This tells that only LHC could produce  the Cho-Maison 
monopole. If so, one might wonder what is the monopole-antimonopole 
pair production rate at LHC. Intuitively the production rate must be 
similar to the WW production, except that the coupling is $4\pi/e$. 
So above the threshold energy, the production rate can be about 
$1/\alpha_{em}$ times bigger than that of the WW production 
rate. 

\section{The MoEDAL Experiment}

The MoEDAL experiment \cite{MoEDAL} is the 7th and latest LHC 
experiment to be approved. The prime purpose of the MoEDAL experiment 
is to search for the avatars of new physics that manifest themselves 
as very highly ionizing particles,  such as the Cho-Maison monopole. 

The MoEDAL experiment will be deployed at Point 8 on the LHC ring in 
the VELO-LHCb cavern. It is due to start data taking in 2015, after 
the long LHC shutdown, when the LHC with be operating at a centre-of-mass 
energy near to 14 TeV. A simplified depiction of the MoEDAL detector is 
shown in Fig. 4.



The mean rate of energy loss per unit length $dE/dx$  of a 
particle  carrying an electric charge $q_{e} = ze$ traveling with velocity $\beta = v/c $  
in a given material is modelled by the Bethe-Bloch formula \cite{bethe}:

\begin{equation}
-\frac{dE}{dx} = K\frac{Z}{A}\frac{q_{e}^{2}}{\beta^{2}}\left[ln\frac{2m_{e}c^{2}\beta^{2}\gamma^{2}}{I} - \beta^{2}\right]
\end{equation}

where $Z, A $ and $I$ are the atomic number, atomic mass and mean excitation energy of the medium,
$K = $ 0.307 MeV g$^{-1}$ cm$^{2}$,  $m_{e}$  is the electron mass and  $\gamma = 1/\sqrt{1 - \beta^{2}}$.
 Higher-order terms are neglected.
 
 For a magnetic monopole carrying a magnetic charge $q_{m} = (ng)ec$, where g is the Dirac charge and $n=1,2,3....$, the velocity
 dependence causes the cancellation of the $1/\beta^{2}$ factor, changing the behaviour of $dE/dx$ at low velocity. 
 The Bethe-Bloch formula becomes:
 
 \begin{figure}[t]
\includegraphics[width=8cm]{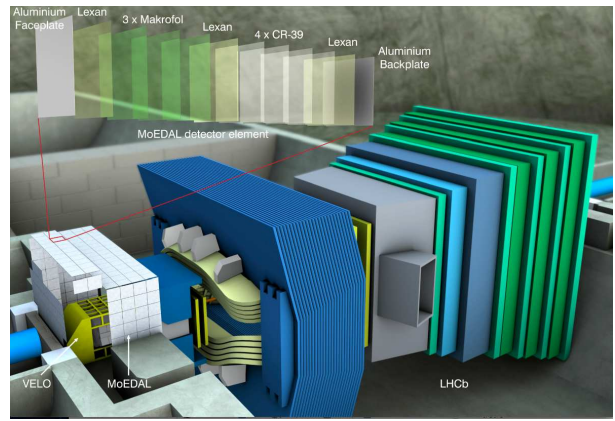}
\caption{A simplified depiction of the MoEDAL detector, adjacent 
to the LHCb detector, at Point 8 on the LHC ring.}
\label{fig2}
\end{figure}
 
\begin{equation}
-\frac{dE}{dx} = K\frac{Z}{A} (ng)^{2}\left[ln\frac{2m_{e}c^{2}\beta^{2}\gamma^{2}}{I_{m}}  + \frac{K(|g|)}{2} -\frac{1}{2} - B(|g|) \right]
\end{equation}

where $I_{m}$  is approximated by the mean excitation energy for electric charges I. The Kazama, Yang
and Goldhaber cross section correction and the Bloch correction are given by $K(|g|)$ = 0.406 (0.346)
for $g$ and $2g$  and $B(|g|)$ = 0.248 (0.672, 1.022, 1.685) for  $g, 2g, 3g, 6g$  \cite{ahlen}, and
are interpolated linearly to intermediate values of $|g|$.  The expression above  is  only valid only down to a 
velocity of $\beta \sim$0.05.

\begin{figure*}
\includegraphics[width=16cm]{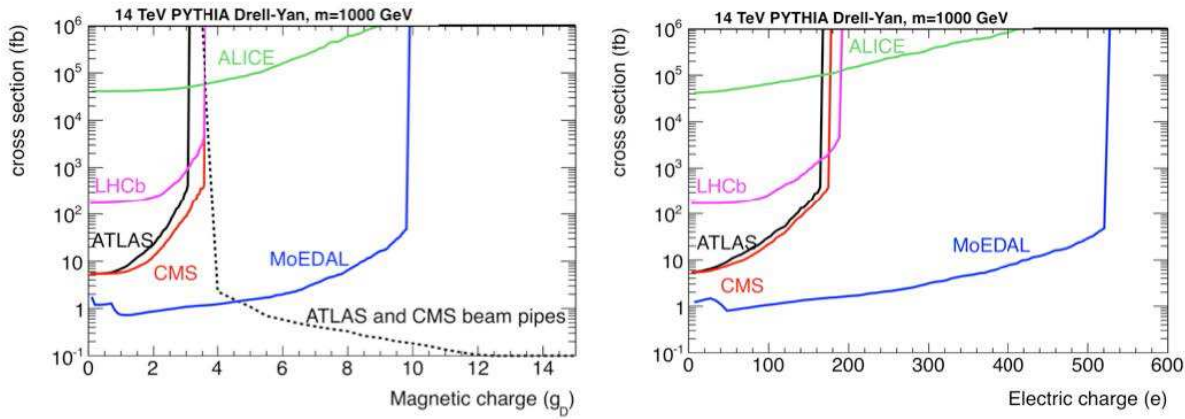}
\caption{The expected reach of the search for the direct detection 
of monopole – anti-monopole pair production produced via the Drell-Yan 
process at the LHC (Ecm =14 TeV). It is assumed here that the
luminosity taken by LHCb/MoEDAL is 2 ${\rm fb}^{-1}$, ATLAS \& CMS 
20 ${\rm fb}^{-1}$, and ALICE only 0.004 ${\rm fb}^{-1}$.}
\label{fig3}
\end{figure*}

One key difference between  a relativistic magnetic monopole with a single Dirac charge
and a electrically charge particle is  that the ionization of a medium caused by the monopole is
very much greater than that of the electrically charged particle. For example, the ionization caused by a 
singly charged, highly relativistic  Dirac monopole is $\sim$4700 times that of, say, a proton moving
 at the same speed. 
 
 However, a singly charged EW monopole has a magnetic charge of $2g$. Thus, the  $dE/dx$ 
 of a highly  relativistic singly charged  EW monopole  is $\sim$ 18,800 (4 x 4700)  times that
 of a highly relativistic monopole. Another important difference between electric and magnetic charge is the behaviour at
low velocity. In the case of electric charge  the dE/dx of the particle increases  with decreasing $\beta$,  giving rise to a large fraction of
the particleÕs energy being deposited near the end of its trajectory  - the so-called Bragg peak.  However, in the case
 of magnetic charge  can be the  dE/dx is expected to $\it diminish$ with decreasing $\beta$.

The  MoEDAL detector consists  of the largest array (over 260 sqm) 
of plastic Nuclear Track Detector (NTD) stacks and trapping detectors 
ever deployed at an accelerator. MoEDAL is a largely passive detector 
which has a dual nature.  First, as a giant camera for "photographing" 
new physics where the NTD systems are the camera's "film" and second 
as a matter trap analyzer for new massive stable magnetically or 
electrically charged particles.

The main general purpose LHC detectors, ATLAS and CMS, are optimized 
for the detection of singly electrically charged particles moving 
near to the speed of light as well as neutral standard model 
particles (neutrons, photons, etc.). On the other hand MoEDAL is 
designed to detect electrically or magnetically charged  particles 
with energy loss greater than or equal to around five times 
that of a minimum ionizing particle. As MoEDAL requires no trigger  
or reader electronics  slowly moving  particles present no problem 
for detection. Thus, the MoEDAL detector operates in a way that 
is complementary to the existing multi-purpose LHC detectors.

A relativistic electroweak monopole has a magnetic charge that 
is a multiple of 2n (n=1,2,3...) that of the Dirac charge - equivalent 
to an ionizing power of ~9400n (n=1,2,3..) times that of a Minimum 
Ionizing Particle (MIP). Thus, it would be rapidly absorbed in the 
beampipe or in the first layers of the massive general purpose LHC  
experiments such as ATLAS and CMS, making it difficult to detect 
and measure. If the fundamental charge was $e/3$ instead of $e$ then 
the ionizing power would be nine time higher making the problem worse.

However, the full width of the MoEDAL plastic NTD detector array 
stacks which are at most only about 5mm thick can be traversed by  
magnetic monopoles with up to around six Dirac charges. In addition, 
the NTD stacks can be calibrated directly for very highly ionizing 
particles using heavy-ion beams, this is not possible with the 
standard LHC experiments. Once calibrated the charge resolution 
of an NTD stack can be as good as 1/100 of a single electric charge. 
A monopole traversing a MoEDAL NTD stack of 10 plastic foils would 
leave a trail of 20 etch pits (allowing a precise measurement of 
the effective charge) aligned with respect to each other to a few 
microns - there is no known standard model background to such a signal.

An additional strength of the MoEDAL detector is the use, for the 
first time at an accelerator, of purpose built trapping detectors 
comprised of aluminium volumes. A fraction of the very highly ionizing 
particles produced during collisions would be trapped in these volumes. 
Periodically, the trapping detectors are replaced and the exposed 
detectors monitored for the presence of trapped magnetic charge 
using a SQUID magnetometer. In this way the MoEDAL detector can 
be used to directly measure the magnetic charge, a first for a 
collider detector. 

Typically the standard LHC detectors need a largish statistical 
sample to establish a signal and measure the basic properties 
of the detected particle. However, the MoEDAL detector requires 
only one particle to be detected in the NTD array and/or captured 
in the trapping detectors in order to declare discovery and to 
determine the basic properties of the particle. In most cases we 
can expect that corroborating evidence should be available from 
the other LHC detectors in the event of a discovery by the MoEDAL 
detector. The reach of the MoEDAL experiment in the physics arena 
where highly ionizing particles are the harbingers of new physics 
is shown in Fig. 5 \cite{moedal-reach}.

\section{Conclusion}

Dirac first hypothesized the existence of the magnetic monopole 
in 1931 \cite{Dirac} as a quantised singularity in the electromagnetic 
field. Since then we have had the Wu-Yang monopole \cite{wuyang}, 
the 'tHooft-Polyakov monopole \cite{thooft}, and the grand unification 
monopole \cite{dokos}, and the quest for magnetic monopoles has 
continued both theoretically and experimentally. But only the electroweak, 
or Cho-Maison, monopole is consistent with the theoretical framework of 
the standard model, where the Dirac monopole becomes the Cho-Maison 
monopole after the electroweak unification. 

The existence of the electroweak monopole invites exciting new 
questions. What is its spin, and how can we predict it?  How can 
we construct the quantum field theory of monopole? What are the new 
physical processes induced by the monopole? What is the impact of the 
monopole on cosmology, particularly the cosmology of the early universe? 

The MoEDAL experiment is optimized to detect very highly ionizing particles
such as the Cho-Maison monopole.  Indeed,  a significant portion
of its  estimated possible mass range  is accessible
at the LHC. If  the Cho-Maison is produced at the LHC with the 
expected cross-section  then  MoEDAL will  detect it.

There is no doubt that the discovery of the Cho-Maison monopole 
would be comparable in importance to  that of the Higgs boson and arguably
more revolutionary. For example, if discovered,  the Cho-Maison  monopole 
will be  the first elementary particle  with magnetic charge and the 
first  elementary particle that is topological in nature.

\end{document}